\documentclass[aps,pre,twocolumn,
,groupedaddress,amsmath,amssymb,amsfonts,showkeys,showpacs]{revtex4}

\usepackage{graphicx}
\usepackage{subfig}

\newcommand{\trademarked}{\textsuperscript{\textregistered}~}

\DeclareMathOperator{\Tr}{tr}

\begin{document}
\title{Constructing Analytically Tractable Ensembles of Non--Stationary Covariances
       with an Application to Financial Data}

\author{Frederik Meudt}
  \email{frederik.meudt@uni-due.de}
  \affiliation{Fakult\"at f\"ur Physik, Universit\"at Duisburg--Essen, Duisburg, Germany}

\author{Martin Theissen}
  \email{martin.theissen@stud.uni-due.de}
  \affiliation{Fakult\"at f\"ur Physik, Universit\"at Duisburg--Essen, Duisburg, Germany}
  
\author{Rudi Sch\"afer}
  \email{rudi.schaefer@uni-due.de}
  \affiliation{Fakult\"at f\"ur Physik, Universit\"at Duisburg--Essen, Duisburg, Germany}
  
\author{Thomas Guhr}
  \email{thomas.guhr@uni-due.de}
  \affiliation{Fakult\"at f\"ur Physik, Universit\"at Duisburg--Essen, Duisburg, Germany}

\date{\today}

\begin{abstract}
  In complex systems, crucial parameters are often subject to
  unpredictable changes in time. Climate, biological evolution and
  networks provide numerous examples for such non--stationarities.  In
  many cases, improved statistical models are urgently called for.  In
  a general setting, we study systems of correlated quantities to which
  we refer as amplitudes. We are interested in the case of
  non--stationarity, \textit{i.e.}, seemingly random covariances. We
  present a general method to derive the distribution of the
  covariances from the distribution of the amplitudes. To ensure
  analytical tractability, we construct a properly deformed Wishart
  ensemble of random matrices. We apply our method to financial
  returns where the wealth of data allows us to carry out
  statistically significant tests.  The ensemble that we find is
  characterized by an algebraic distribution which improves the
  understanding of large events.
\end{abstract}

\pacs{05.10.-a, 05.40.-a, 05.45.-a, 89.65.Gh, 89.75.-k}

\keywords{non--stationarity, covariances, random matrices, compounding, financial data}

\maketitle

\section{Introduction}

In electroencephalography (EEG) electrical currents are recorded at
different positions on the scalp to measure the brain activity. The
correlations between the time series of these currents strongly depend
on the overall state of the brain.  During an epileptic seizure, for
example, the correlations are much stronger than in normal
periods~\cite{pijn1991chaos, muller2005detection}. This time dependence of the correlations is the
kind of non--stationarity that we wish to address.
Non--stationarities are also seen when wave packets travel through
disordered systems. Even if the disorder is static, the correlations
between the wave intensities measured at different
positions versus time will change, when the direction or the composition of the
wave packet is altered~\cite{hohmann2010freak, metzger2014statistics, degueldre2015random}. Finance provides
another important example for this type of non--stationarity.  The
correlations between stock price time series change in time, just as
the business relations between the firms and the traders' market
expectations~\cite{bekaert1995time,Longin1995,Onnela2003,Zhang2011,Sandoval2012,muennix2012}.  Similar
non--stationarities exist in many complex systems, including velocity
fluctuations in turbulent flows, heartbeat dynamics, series of waiting
times, etc. \cite{Ghasemi2008, PhysRevE.87.062139, PhysRevE.75.060102,
  Schafer20103856}.

A system showing non--stationary correlations may be interpreted as
being out of equilibrium, implying that some of the key tools in
statistical physics are not applicable. Yet, the challenges are
similar to the one faced for equilibrium systems: Is there generic or
universal behavior? --- How can we identify it? -- Can we set up
statistical models for these non--stationarities? --- In the context
of finance, we recently put forward a random matrix approach to tackle
these issues~\cite{Schmitt2013}. We also succesfully applied it in a
study of credit risk and its impact on systemic
stability~\cite{schmitt2014credit}. Inspite of the conceptual differences,
random matrix theory~\cite{meh04,guhr98} formally has much in common
with statistical mechanics. Observables are averaged over an ensemble;
in statistical mechanics, it usually is the microcanonical, canonical
or macrocanonical one, in random matrix models, it is an ensemble of those
matrices which describe or characterize the system.  In the context of
the present discussion, random matrix models can be divided into two
classes:
\begin{enumerate}\setlength{\itemsep}{-2pt}
\item The ensemble is fictitious. It comes into play via an ergodicity
  argument only.
\item The ensemble really exists and can be identified in the
  system. The issue of ergodicity does not arise.
\end{enumerate}
The vast majority of random matrix models in, \textit{e.g.}, quantum
chaos falls into class 1, for a review see Ref.~\cite{guhr98}. One is
interested in the spectral statistics of one individual system. Its
Hamiltonian is viewed as a random matrix, whose dimension is
eventually sent to infinity. Ergodicity holds in this limit, meaning
that a smoothing energy average of an observable over one individual
spectrum equals the average over an ensemble of random matrices. A
noticeable exception are random matrix applications to quantum
chromodynamics~\cite{Jac00}. In lattice gauge theory, the quarks first
propagate in frozen configurations of the gauge fields, before an
average over the gauge fields, modeled by random matrices, is carried
out as second step. This clearly belongs in class 2. The fluctuating
gauge fields truly exist, the partition function involves an integral
over them. Ergodicity reasoning is not evoked.

There are numerous applications of random matrix theory in
finance~\cite{lal99,lal99b,laloux00,ple99,ple02,pafka04,potters2005financial,drozdz2008empirics,kwapien2006bulk,biroli2007student,burda2011applying} which address statistical properties of 
correlation matrices. Many of them also deal with non--Gaussian
ensembles. To the best of our knowledge, all of these applications
fall into class 1, because one is interested in the statistics of one
individual correlation matrix, measured at one particular instant in
time.  In our study~\cite{Schmitt2013}, we put forward a first
application of random matrices in finance that belongs in class 2.
Non--stationarity makes the covariances fluctuate and thereby creates
an ensemble of covariance matrices which we approximated by a Gaussian
Wishart ensemble of random matrices~\cite{Wishart1928}. We derived how
the multivariate distribution of dimensionless price changes, referred
to as returns, acquires heavy tails due to the
non--stationarity. Hence, we showed that the non--stationarities
indeed have universal features.

Here, we have three goals: First, we present in Sec.~\ref{sec2} a
statistically significant way to construct a proper and analytically
tractable random matrix ensemble from the data. We emphasize that this
is an important issue for random matrix models in the context of
correlations. In contrast to quantum chaos, where universality holds
on the scale of the mean level spacing, there is not such a local
scale when studying statistical properties of correlation
matrices. Thus, a Gaussian assumption is not always justified and it
does matter what the ensemble looks like in reality. In particular,
realistic ensembles considerably help to understand and model large
events. Our construction is general and not tied to any specific
system. Its merit lies in the fact that once the enemble is known, it
can be used to work out generic statistical properties of any
observable depending on the correlation matrices, see
Ref.~\cite{schmitt2014credit} for an example.  Second, we apply our approach
to financial data in Sec.~\ref{sec3}. We identify an algebraic
ensemble, which is quite relevant for risk estimation. Third, we
discuss two issues arising in our general construction in
Secs.~\ref{sec4} and~\ref{sec5}, namely a certain conceptual caveat
and yet a further extension, respectively. Conclusions are given in
Sec.~\ref{sec6}.

\section{Constructing a Proper Random Matrix Ensemble}
\label{sec2}

After setting up the general problem in Sec.~\ref{sec21}, we introduce
the deformed Wishart ensemble and derive the correponding amplitude
distribution in Sec.~\ref{sec22}. The determination of the deformation
functions which characterize the ensemble and the amplitude
distribution is discussed in Sec.~\ref{sec23}. Here, we derive the
approach for the general case, for sake of illustration, the reader is
referred to Ref.~\cite{Schmitt2013} and Sec.~\ref{sec3}

\subsection{Non--Stationary Covariances}
\label{sec21}

Suppose we have measured in a system with randomness $K$ amplitudes as
time series $R_k(t), k=1,\ldots,K$ over a long interval
$T_\textrm{tot}$ of time $t=1,\ldots,T_\textrm{tot}$.  For examples,
these amplitudes can be electric or magnetic fields at $K$ different
points in a disordered system, positions of $K$ randomly moving
particles or financial returns, \textit{i.e.} dimensionless price
changes for $K$ stocks. Importantly, we assume that there are
correlations between the time series. In complex systems, one often
encounters the situation that crucial system parameters, in particular
the covariances or correlations, are seemingly random functions of
time~\cite{ball2000stochastic,burtschell2005beyond,ankirchner2012cross}. 
To be more precise, we consider a time window of
length $T$ that is much shorter than the total interval, $T\ll
T_\textrm{tot}$. We now want to average over the subinterval $[t-T+1,t]$ of 
length $T$ whose position in the total interval is determined by the
time $t$. Sample averages of a function $f(t)$ in this subinterval
are then written as
\begin{equation}
\langle R_k \rangle_T(t) =  \frac{1}{T} \sum_{t'=t-T+1}^t f(t') \ .
 \label{returns}
\end{equation}
We are particularly interested in the covariances
\begin{eqnarray}
 \Sigma_{kl}(t) &=& \langle R_k R_l\rangle_T(t) - 
               \langle R_k\rangle_T(t) \langle R_l\rangle_T(t) \nonumber\\
            &=& \langle r_k r_l\rangle_T(t) \ ,
\label{covar}
\end{eqnarray}
where we introduced the amplitudes normalized to zero mean value,
\begin{equation}
 r_k(t) = R_k(t)-\langle R_k \rangle_T(t) \ .
 \label{returns}
\end{equation}
We keep in mind that the resulting $K\times K$ covariance matrix
$\Sigma(t)$ is calculated from time series of length $T$. We now move
this time window of length $T$ through the data, the resulting
covariances $\Sigma_{kl}(t)$ fluctuate. This non--stationarity has an
important impact on other statistical observables. In the present
study, we focus on the distribution of the amplitudes. We now consider
a time interval $T$ as short as possible such that the covariance
matrix $\Sigma_s$ in this time interval is in good approximation
constant. We begin with addressing the case in which the distribution
of the amplitudes is, for a given time $t$, well approximated by a
multivariate Gaussian
\begin{equation}
 g(r|\Sigma_s) = \frac{1}{\sqrt{\det{\left(2\pi\Sigma_s\right)}}}
                     \exp{\left(-\frac{1}{2}r^\dagger \Sigma_s^{-1} r \right)} 
\label{gaussian}
\end{equation}
with the $K$ component vector $r=(r_1,\ldots,r_k)$ and the $K\times K$
covariance matrix $\Sigma_s$. We suppress the argument $t$ of $r$ and
use $\dagger$ to indicate the transpose. We refer to $g(r|\Sigma_s)$
as \textit{static amplitude distribution}.  Due to the correlations, a
Gaussian assumption for the static distribution is not as restrictive
as it may seem. In the eigenbasis of $\Sigma_s$, the amplitudes only
appear in linear combinations.  Thus, for large $K$, the mechanisms
that lead to the central limit theorem start working and drive the
distributions towards Gaussians.  Later on in Sec.~\ref{sec5} we will
nevertheless relax the Gaussian assumption for the static amplitude
distribution and look at more general functional forms.

\subsection{Deformed Wishart Ensemble and its Amplitude Distribution}
\label{sec22}

How does the non--stationarity affect the amplitude distribution when
data from the total interval $T_\textrm{tot}$ are analyzed? --- As in
Ref.~\cite{Schmitt2013}, we model this by random matrices. As the
covariance matrix is different at each time $t$ where it is analyzed,
we replace the covariance matrix in the distribution (\ref{gaussian})
by the expression
\begin{equation}
 \Sigma_s \longrightarrow \frac{1}{N}AA^\dagger \ ,
\label{repl}
\end{equation}
where $A$ is a real rectangular $K\times N$ random matrix without any
symmetries. The right hand side of Eq.~(\ref{repl}) has to have the
form given to ensure that it can model a properly defined covariance
matrix. This follows directly from the definition~(\ref{covar}).
Although $K$, the first dimension is fixed, the second one, $N$, is
for the time being a free model parameter. It can be viewed as the
length of the model time series. Further clarifications will follow.
To obtain the amplitude distribution for the total interval, we average
over the random matrices 
\begin{equation}
 \langle g\rangle (r|\Sigma,N) = \int d[A] \overline{w}(A|\Sigma,N) 
                    g\left(r\left|\frac{1}{N}AA^\dagger\right.\right) \ ,
\label{average}
\end{equation}
where $d[A]$ is the volume element, \textit{i.e.}, the product of all
independent variables in $A$. Following
Wishart~\cite{Wishart1928,Muirhead1982}, the Gaussian distribution
\begin{equation}
w(A|\Sigma) = \frac{1}{\det^{N/2}{\left(2\pi\Sigma\right)}}
          \exp{\left(-\frac{1}{2}\Tr A^\dagger\Sigma^{-1}A\right)}
\label{uswis}
\end{equation}
was assumed for the random matrices in Ref.~\cite{Schmitt2013}.  It
describes the Gaussian fluctuations of the model covariance matrices
$AA^\dagger/N$ about the given empirical covariance matrix $\Sigma$,
which is evaluated over the total time interval of length
$T_\textrm{tot}$. It should not be confused with the empirical
covariance matrix $\Sigma(t)$ calculated in the subintervals
$[t-T+1,t]$.  The crucial difference compared to
Ref.~\cite{Schmitt2013} is a generalization of the
ensemble~(\ref{uswis}).  We introduce the deformed Wishart ensemble
\begin{equation}
\overline{w}(A|\Sigma,N) = \int\limits_0^{\infty}d\eta f(\eta)
                      w\left(A\left|\frac{N\Sigma}{\eta}\right.\right)
\label{defwis}
\end{equation}
which is defined by the \textit{ensemble deformation function}
$f(\eta)$ with the properties
\begin{align}
\int\limits_0^\infty f(\eta) d\eta = 1 \quad \textrm{and} \quad f(\eta) \ge 0 \ .
\label{fconstraints}
\end{align}
For later convenience, $\Sigma$ on the right hand side of
Eq.~(\ref{defwis}) is rescaled with $N$.  The flutuations of the model
covariance matrices $AA^\dagger/N$ deviate from Gaussian, but
always about the empirical covariance matrix $\Sigma$.  The meaning of
the model parameter $N$ now becomes clearer.  It sets the variance for
these fluctuations. The above rescaling only changes the functional
dependencies, but not the r\^ole of $N$. We emphasize once more that
$\Sigma$ is evaluated over the total time interval. Similar
deformations of random matrix ensembles but in a Hamiltonian, not
Wishart setting were apparently first put forward in
Refs.~\cite{Bertuola2004,MuttalibKlauder2005}.

After inserting the ansatz~(\ref{defwis}) into Eq.~(\ref{average}),
we may use the result~\cite{Schmitt2013}
\begin{equation}
\int w(A|\Sigma) g\left(r\left|\frac{1}{N}AA^\dagger\right.\right) d[A]
  = \int\limits_0^\infty \chi^2_N(z) g\left(r\left|\frac{z}{N}\Sigma\right.\right) dz \ ,
\label{gaver}
\end{equation}
which reformulates the whole random matrix average as a univariate
average over the $\chi^2$ distribution
\begin{equation}
\chi_N^2(z) = \frac{1}{2^{N/2}\Gamma(N/2)} z^{N/2-1} \exp\left(-\frac{z}{2}\right)
\label{chi2}
\end{equation}
of $N$ degrees of freedom. On the mathematical side, there are
connections between formula~(\ref{gaver}) and the calculation of
certain distributions in scattering theory~\cite{Poli09,Yan12}.  Using
the result~(\ref{gaver}), the amplitude distribution reduces to the
double integral
\begin{equation}
 \langle g\rangle (r|\Sigma,N) = \int\limits_0^{\infty}d\eta f(\eta)
                                 \int\limits_0^\infty dz \chi^2_N(z)
                    g\left(r\left|\frac{z}{\eta}\Sigma\right.\right) \ .
\label{averagedouble}
\end{equation}
Again, we point out the rescaling of $\Sigma$ with $N$, \textit{cf.}
Eq.~(\ref{gaver}). It is useful to rewrite that as a single integral
\begin{equation}
 \langle g\rangle (r|\Sigma,N) = \int\limits_0^{\infty} p(x)
                    g\left(r|x\Sigma\right) dx
\label{averagesingle}
\end{equation}
by introducing the variable $x=z/\eta$ and its distribution 
\begin{equation}
p(x) = \int\limits_0^{\infty}d\eta f(\eta)
              \int\limits_0^\infty dz \chi^2_N(z) \delta\left(x-\frac{z}{\eta}\right) \ .
\label{aampldef}
\end{equation}
We refer to it as \textit{amplitude distribution deformation function}. One easily
obtains
\begin{equation}
p(x) = \frac{x^{N/2-1}}{2^{N/2}\Gamma(N/2)}
              \int\limits_0^{\infty}d\eta f(\eta) \eta^{N/2}\exp\left(-\frac{\eta x}{2}\right) \ ,
\label{pdis}
\end{equation}
which establishes the relation between the two deformation functions.

We notice that the ansatz~(\ref{defwis}) restricts the form of the
deformed distribution $\overline{w}(A|\Sigma,N)$ to functions of $\Tr
A^\dagger\Sigma^{-1}A$ only. Even though the inclusion of further
terms such as $\Tr(A^\dagger\Sigma^{-1}A)^2$ is likely to improve the
quality of the data fits, we stick to the ansatz~(\ref{defwis}). Its
considerable advantage is the guaranteed but otherwise questionable
analytical tractability as will be shown in the sequel. Moreover,
further terms will also increase the number of deformation functions
which will hamper their unambiguous determination.

\subsection{Determination of the Deformation Functions}
\label{sec23}

Apart from the deformation functions, the distributions
$\overline{w}(A|N\Sigma)$ and $\langle g\rangle (r|\Sigma,N)$ depend
on the usual covariance matrix $\Sigma$ analyzed by sampling over the
total interval. We notice that the corresponding covariance matrix
$\Sigma^{(d)}$ in the deformed ensemble slightly differs from that.
By definition we have
\begin{eqnarray}
\Sigma^{(d)} &=& \langle \frac{1}{N} AA^\dagger \rangle \nonumber\\
             &=& \int \frac{1}{N}AA^\dagger \overline{w}(A|\Sigma,N) d[A] \ .
\label{sigmad}
\end{eqnarray}
Inserting Eq.~(\ref{defwis}), we can do the ensemble average in the
Gaussian case which yields the covariance matrix $N\Sigma/\eta$. Thus,
only the $\eta$ integral remains and we have
\begin{equation}
\Sigma^{(d)} = N \Sigma \int\limits_0^\infty \frac{f(\eta)}{\eta} d\eta
             = N \Sigma \overline{\eta^{-1}} 
\label{sigmadd}
\end{equation}
implying that the two covariance matrices differ by the average of $1/\eta$.
Alternatively, one can calculate $\Sigma^{(d)}$ from the amplitude distribution,
\begin{equation}
\Sigma^{(d)} = \langle rr^\dagger \rangle
             = \int rr^\dagger \langle g\rangle (r|\Sigma,N) d[r] \ ,
\label{sigmada}
\end{equation}
which yields
\begin{equation}
\Sigma^{(d)} = \Sigma \int\limits_0^\infty x p(x) dx
             = \Sigma \overline{x} \ . 
\label{sigmadad}
\end{equation}
Here, the two covariance matrices differ by the first moment of $x$.
With the help of Eq.~(\ref{aampldef}), the
results~(\ref{sigmadd},\ref{sigmadad}) are easily seen to coincide.

Having extracted the covariance matrix for the total time interval
from the data, we can proceed with the determination of the
deformation functions.  The exponential function in the integrand of
Eq.~(\ref{pdis}) allow us to interpret it as a Laplace transform,
\begin{equation}
\frac{\Gamma(N/2)}{2} \frac{p(x)}{x^{N/2-1}}
 = \mathcal{L}\left(\tilde{\eta}^{N/2}f(2\tilde{\eta})\right) \ ,
 \label{laplace}
\end{equation}
where we introduced $\tilde{\eta}=\eta/2$ to avoid inconvenient factors
of two. Thus, the ensemble deformation function is the inverse
Laplace transform
\begin{equation}
f(2\tilde{\eta}) = \frac{\Gamma(N/2)}{2}\frac{1}{\tilde{\eta}^{N/2}}
                   \mathcal{L}^{-1}\left(\frac{p(x)}{x^{N/2-1}}\right) \ .
 \label{invlaplace}
\end{equation}
of the amplitude distribution deformation function divided by a power
of $x$. This makes it possible to determine $f(\eta)$ by extracting
$p(x)$ from the amplitude time series and carrying out the inverse
Laplace transform.  In contrast, extracting $f(\eta)$ directly from
the data is cumbersome and burdened by limited statistics, as the
following discussion shows. The rows of $A$ are the model time series
of length $N$ and cannot easily be identified with the amplitude time
series $r_k(t)$ of lenght $T$. However, the matrices $AA^\dagger/N$
form the ensemble of model covariance matrices and can be compared
with the empirical ones.  As a certain sample length is required for
meaningful results, it is out of question to compare the matrices
directly, \textit{i.e.}  their individual matrix elements. A better
observable is the distribution
\begin{equation}
q(s) = \int \delta\left(s-\frac{1}{N}\Tr AA^\dagger\right)\overline{w}(A|\Sigma,N) d[A]
\label{trace}
\end{equation}
of the traces, which can easily be written as a single integral
involving the ensemble deformation function $f(\eta)$. The
distribution~(\ref{trace}) is empirically obtained by moving a time
window through the amplitude time series and calculating the empirical
covariance matrices and their traces. This then gives $f(\eta)$.

The problem with the above procedure is its still limited statistical
significance. Instead, extracting the amplitude distribution
deformation function $p(x)$ from the data gives much more meaningful
results. As we discuss in the sequel, the number of data points is
larger by a factor of $K$. The amplitudes $r_k$ appear in
Eq.~(\ref{averagesingle}) only via the bilinear form $r^\dagger\Sigma
r$. We rotate the amplitude vector $r$ into the eigenbasis of the
empirically obtained covariance matrix $\Sigma$.  By definition, the
eigenvalues of $\Sigma$ are positive and larger than zero, provided
the length of the sampling interval is larger than $K$. We divide each
component of the rotated amplitude vector by the square root of the
corresponding eigenvalue and denote the resulting vector by
$\tilde{r}$.  Within our model, all components of $\tilde{r}$ should
be equally distributed. We integrate out all but one, $\tilde{r}_k$,
and arrive at the distribution
\begin{equation}
 \langle \tilde{g}\rangle (\tilde{r}_k) = \int\limits_0^{\infty} p(x)
     \frac{1}{\sqrt{2\pi x}}\exp\left(-\frac{\tilde{r}_k^2}{2x}\right) dx \ .
\label{rotscale}
\end{equation}
Thus, $p(x)$ may be identified with the distribution of the variances
$x$ of the Gaussian distributed random variables
$\tilde{r}_k$. Conceptually, this is our main result.  It provides a
simple and statistically significant method to obtain the amplitude
distribution deformation function $p(x)$ which then yields upon
inverse Laplace transform the ensemble distribution function
$f(\eta)$. As we have $K$ time series $r_k(t)$, we gain a factor of
$K$ by aggregation.

\section{Application to Financial Data}
\label{sec3}

We now apply our method to stock market data. This is of particular
interest as heavy tails are ubiquituous in finance. A better modeling
for multivariate distributions is urgently called for to improve risk
estimation. We present the data in Sec.~\ref{sec31}, extract the
deformation functions in Sec.~\ref{sec32} and calculate the ensemble
and return distributions in Sec.~\ref{sec33}.

\subsection{Data Set}
\label{sec31}

\begin{figure}[htbp]
  \begin{center}
    \includegraphics[width=0.235\textwidth]{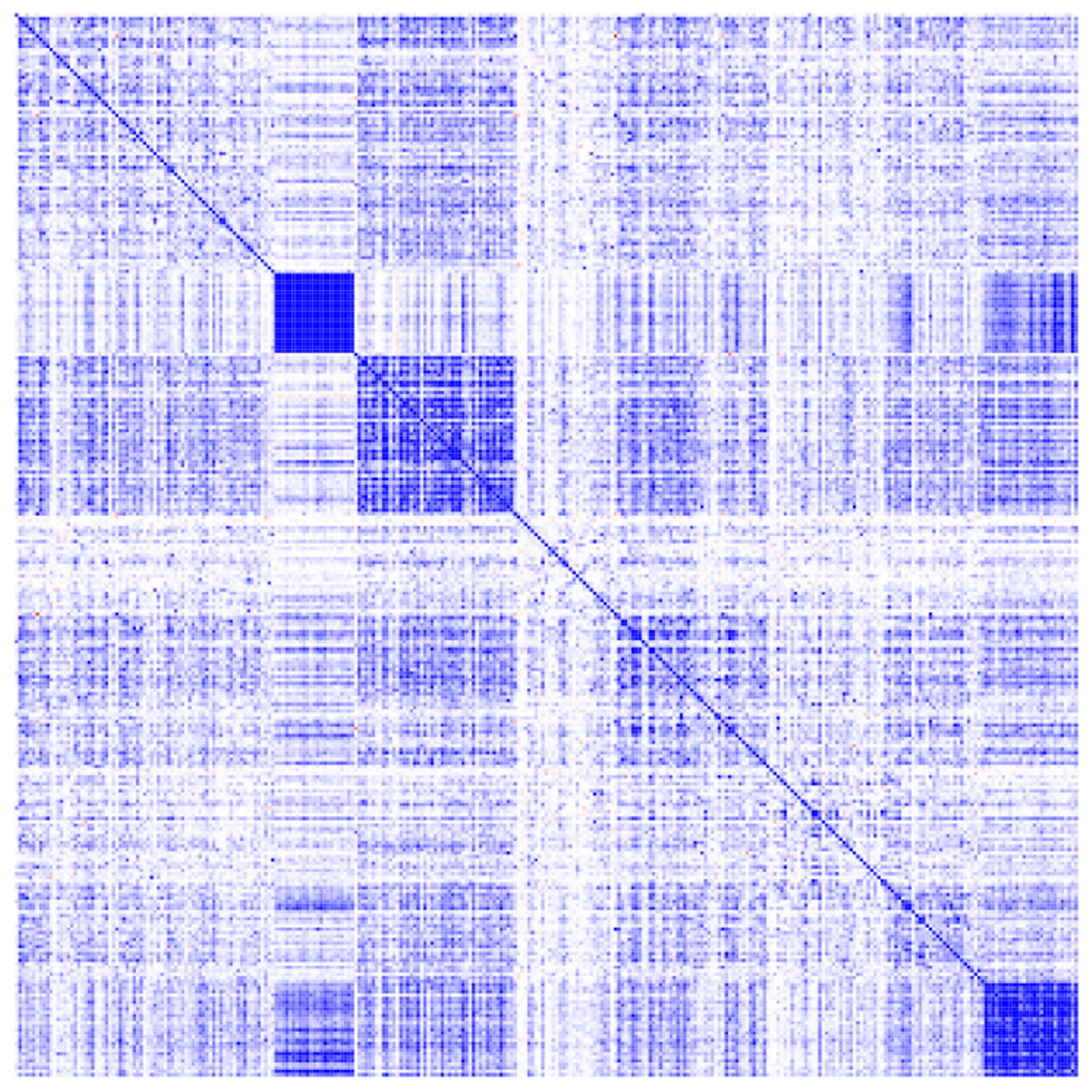}
    \includegraphics[width=0.235\textwidth]{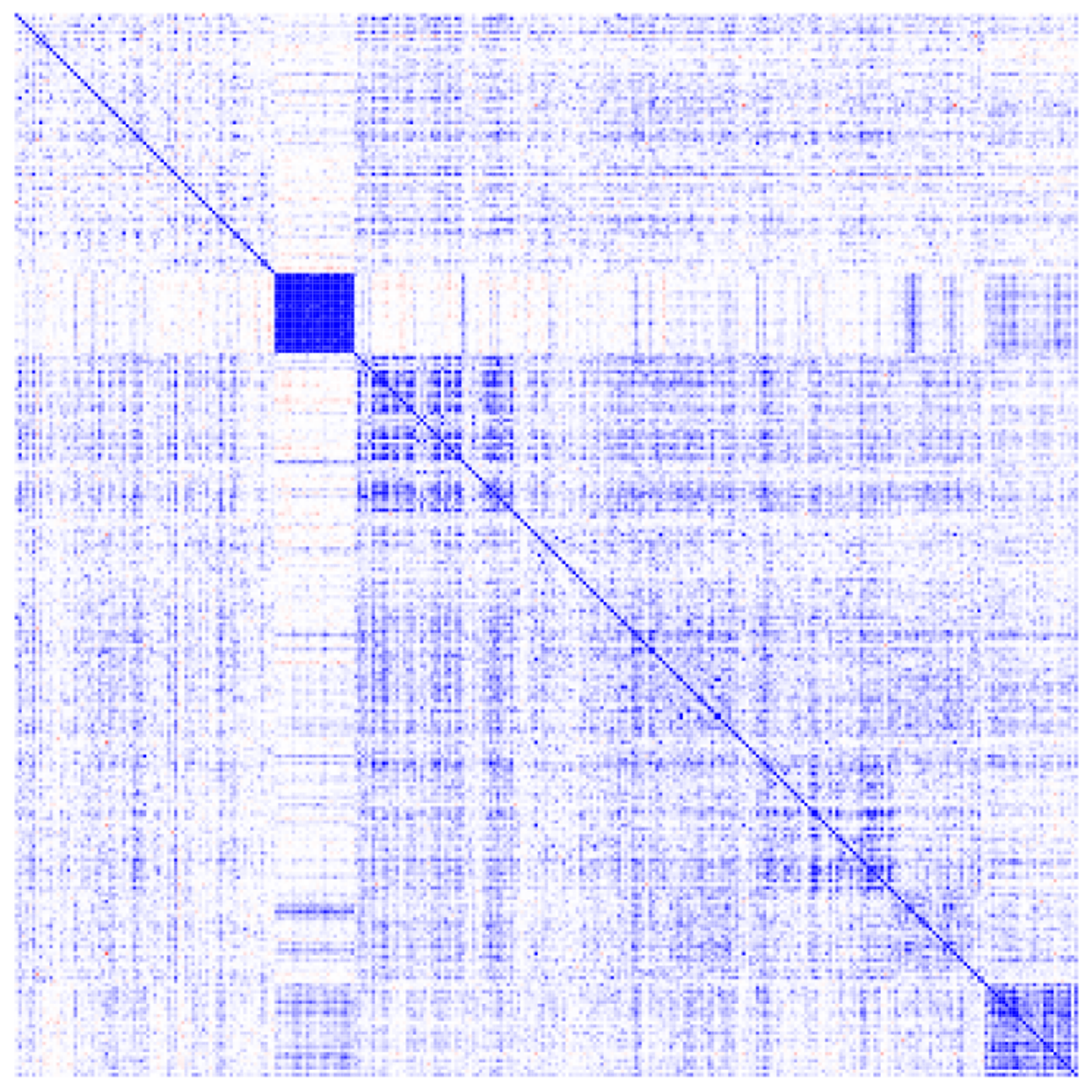}
  \end{center}
  \caption{Correlation matrices of $K=306$ companies in S\&P
    500\trademarked for the fourth quarter of 2005 and the first
    quarter of 2006. The darker, the stronger the correlation. The
    companies are sorted according to industrial sectors. Taken from
    Ref.~\cite{Schmitt2013}.}
 \label{fig1}
\end{figure}

We analyze the $K=306$ continuously traded stocks with
prices $S_k(t), \ k=1,\ldots,K$ in the S\&P 500\trademarked index
between 1992 and 2012~\cite{yahoo}, which we previously analyzed with
a purely Gaussian, \textit{i.e.}, non--deformed Wishart
ensemble~\cite{Schmitt2013}. The amplitudes are here the dimensionless
price changes
\begin{equation}
 R_k(t) = \frac{S_k(t+\Delta t)-S_k(t)}{S_k(t)} \ ,
 \label{returng}
\end{equation}
which are referred to as returns. They depend on the chosen return
horizon $\Delta t$. According to Eq.~(\ref{returns}), we calculate the
returns $r_k(t)$ normalized to zero mean.  To make our presentation
self--contained, we show once more how strongly the whole $K\times K$
correlation matrix $C$ for this data set changes in time.  In
Fig.~\ref{fig1}, it is displayed for subsequent three--months time
windows. In most random matrix approaches, the ensemble is fictitious
and enters only by means of an ergodicity argument. This is not so
here, as Fig.~\ref{fig1} illustrates. Our ensemble exists in reality,
it is the whole set of matrices analyzed by moving a sample time
window through the data. In Fig.~\ref{fig1}, one also sees rather
stable stripes in these correlation matrices which are due to the
different industrial sectors, see, \textit{e.g.},
Ref.~\cite{muennix2012}. Obviously, basis invariance is not present in
this data set, and probably neither in any other real data set. Hence,
a direct extraction of the ensemble deformation function $f(\eta)$
which preserves the basis invariance of the random matrix ensemble is
problematic.  Yet, there is still another reason: market states were
identified which reveal a fine structure of the
ensemble~\cite{muennix2012,chetalova2015zooming}. As every random matrix
ensemble has an effective character, one is advised to analyze
quantities which already reflect this. In the present case, such
quantities are the amplitude, in the present case return, distribution
and the corresponding deformation function $p(x)$.

\subsection{Deformation Functions}
\label{sec32}

\begin{figure}[htbp]
  \begin{center}
    \includegraphics[width=0.4\textwidth]{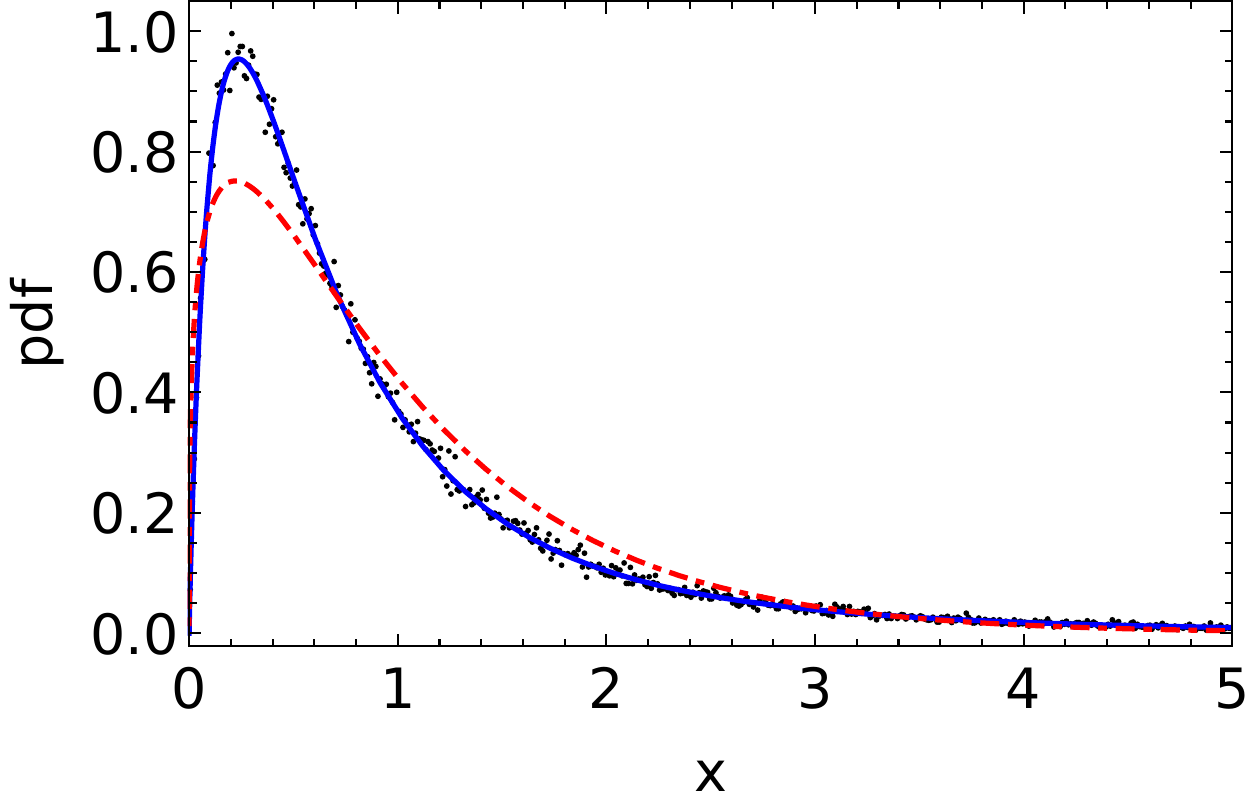}
  \end{center}
  \begin{center}
    \includegraphics[width=0.4\textwidth]{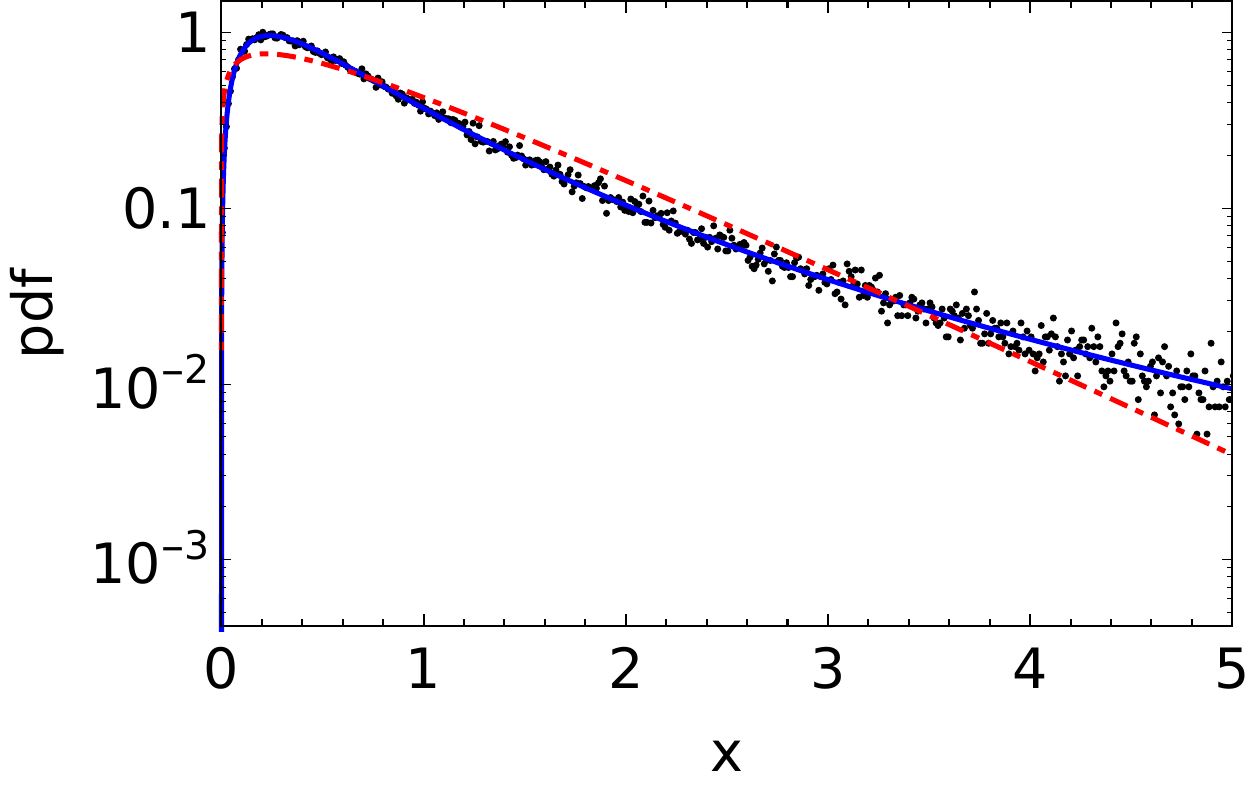}
  \end{center}
  \caption{Aggregated distribution of variances as dots, calculated
    on a five--day time window on a linear (top) and logaritmic
    (bottom) scale. Fit to a beta prime distribution as solid
    line. For comparison, a $\chi^2$ distribution as dashed dotted
    line.}
 \label{fig2}
\end{figure}

\begin{figure}[htbp]
  \begin{center}
    \includegraphics[width=0.4\textwidth]{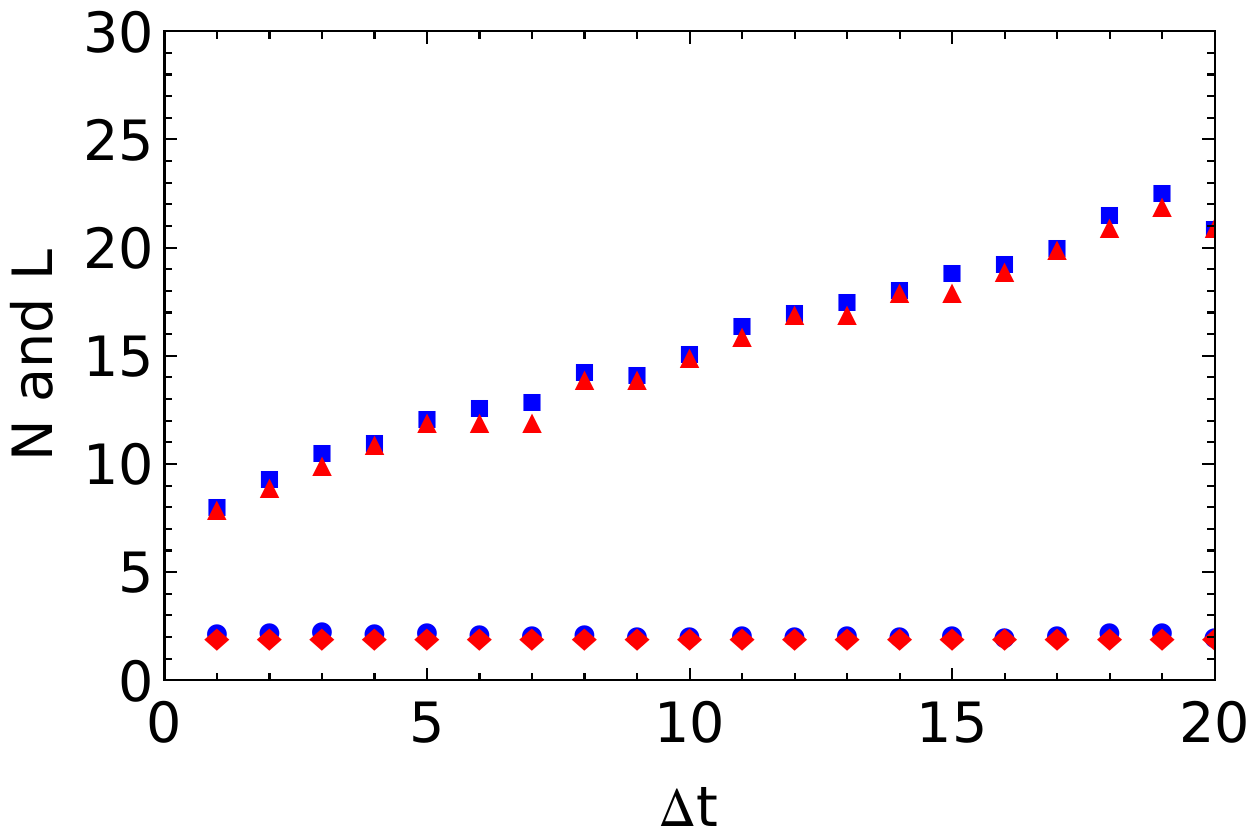}
  \end{center}
  \caption{Fitted values of $N$ (upper, ascending dots) and $L$
    (lower dots) for the beta prime distribution versus the return
    horizon $\Delta t$. Triangles and diamonds for the fits with
    constraint to integer values, suares and circles without
    constraint.}
 \label{fig3}
\end{figure}

We use daily data, \textit{i.e.}, $\Delta t=1$ trading day. Rotation
of the return vector $r$ into the eigenbasis of the empirical
covariance matrix $\Sigma$, normalization to the square roots of the
eigenvalues and aggregation on a five--day window yield the empirical
distributions of variances shown in Fig.~\ref{fig2}. Aggregation on a
ten--day window gives similar results; hence, the estimation noise does not have a major impact on the distribution. A variety of functions is
capable of describing the data.  In finance, one often employs the
log--normal distributions to model volatilities, see \textit{e.g.}
Ref.~\cite{Micciche2002756}. In finance, the standard deviations are
referred to as volatilities.  However, the log--normal distribution
fails to capture the empirically found tail behavior.  More suitable is
the beta prime distribution
\begin{equation}
  p(x|N,L) = \frac{\Gamma(N+L/2)}
                  {\Gamma(N/2)\Gamma((N+L)/2)} \frac{x^{N/2-1}}{(1+x)^{N+L/2}}
 \label{betaprime}
\end{equation}
with two positive parameters $N$ and $L$. Anticipating the following
discussion, we choose their combination in the
expression~(\ref{betaprime}) in such a way that $N$ coincides with the
parameter $N$ introduced in Sec.~\ref{sec2}. The fit is depicted in
Fig.~\ref{fig2}, the agreement with the data is much better than for a
$\chi^2$ distribution corresponding to the ensemble of
Ref.~\cite{Schmitt2013} which is formally obtained by setting
$f(\eta)=\delta(\eta-1)$ or $f(\eta)=\delta(\eta-N)$, respectively,
depending on the rescaling with $N$. We carry out fits for different
return horizons $\Delta t$. The results for $N$ and $L$ are shown in
Fig.~\ref{fig3}. While $L$ stays constant around two, $N$ increases
from about seven for daily data to about 23 for $\Delta t=19$ trading
days. We postpone the interpretation up to the evaluation of the
ensemble and return distribution.

Having extracted the amplitude, \textit{i.e.}, return distribution
deformation function $p(x|N,L)$, we calculate the inverse Laplace
transform~(\ref{invlaplace}),
\begin{equation}
  f(2\tilde{\eta}|N,L) = 
         \frac{\Gamma(N+L/2)}{2\tilde{\eta}^{N/2}\Gamma((N+L)/2)}
               \mathcal{L}^{-1}\left((1+x)^{-N-L/2}\right) \ ,
\label{betaprime1}
\end{equation}
and find~\cite{Gradshteyn2007} with $\eta=2\tilde{\eta}$ for
the ensemble deformation function
\begin{eqnarray}
f(\eta|N,L) &=& \frac{\eta^{(N+L)/2-1}}{2^{(N+L)/2}\Gamma((N+L)/2)}
                   \exp{\left(-\frac{\eta}{2}\right)} \nonumber\\
            &=& \chi_{N+L}^{2}(\eta) \ .
\label{betaprime2}
\end{eqnarray}
This is a $\chi^2_{N+L}$ distribution with $N+L$ degrees of
freedom. As required, $f(\eta|N,L)$ is a positive and normalized
function.

\subsection{Deformed Ensemble and Return Distribution}
\label{sec33}

Inserting Eq.~(\ref{betaprime2}) into Eq.~(\ref{defwis}) yields
after a straightforward calculation  
\begin{eqnarray}
\overline{w}(A|\Sigma,N,L) &=& \frac{\Gamma((N+NK+L)/2)}
                                    {\Gamma((N+L)/2)\det^{N/2}(\pi N\Sigma)} \nonumber\\
                 & & \left(1+\frac{\Tr A^\dagger\Sigma^{-1}A}{N}\right)^{-(N+NK+L)/2} \ .
\label{wisal}
\end{eqnarray}
for the distribution of the random matrices $A$. Thus, we arrive at an
ensemble characterized by an algebraic distribution. For a similar
ensemble, but in the special case of $\Sigma=1_K$, spectral
correlation functions were studied in
Refs.~\cite{Akemann2008,Abul-Magd2009}. Here, however, we derived our
ensemble from data, and the dependence on a non--trivial $\Sigma$ is
in the present essential. Anticipating the result~(\ref{wisal}), we
rescaled $\Sigma$ with $N$ as compared to Ref.~\cite{Schmitt2013}.
Thereby, $N$ and $L$ appear on equal footing in the formulae.  To
obtain the ensemble averaged return distribution, we plug
Eq.~(\ref{betaprime}) into Eq.~(\ref{averagesingle}) and find
\begin{eqnarray}
& & \langle g\rangle (r|\Sigma,N,L) = \frac{\Gamma(N+L/2)\Gamma((N+K+L)/2)}
                                    {\Gamma(N/2)\Gamma((N+L)/2)\sqrt{\det(\pi N\Sigma)}} \nonumber\\
& & \qquad \mathcal{U}\left(\frac{N+K+L}{2},\frac{K-N+2}{2},\frac{r^\dagger\Sigma^{-1}r}{2}\right) 
\label{wisret}
\end{eqnarray}
with the confluent hypergeometric function~\cite{Abramowitz1972}
\begin{equation}
 \mathcal{U}(a,b,z) = \frac{1}{\Gamma(a)}\int\limits_0^\infty y^{a-1} (1+y)^{b-a-1} \exp(-yz) dy
\label{hyper}
\end{equation}
for positive real parts of $a$ and $z$. From Eq.~(\ref{sigmadd}) or (\ref{sigmadad})
the covariance matrix
\begin{equation}
\Sigma^{(d)} = \frac{N}{N+L-2} \Sigma 
\label{covard}
\end{equation}
for the deformed ensemble follows.  To compare with the empirical
return distribution, we compute the integral~(\ref{rotscale}),
\begin{eqnarray}
& & \langle \tilde{g}\rangle (\tilde{r}_k|N,L) = \frac{\Gamma(N+L/2)\Gamma((N+L+1)/2)}
                                    {\Gamma(N/2)\Gamma((N+L)/2)\sqrt{2\pi}} \nonumber\\
& & \qquad\qquad \mathcal{U}\left(\frac{N+L+1}{2},\frac{3-N}{2},\frac{\tilde{r}_k^2}{2}\right) \ .
\label{wisretmarginal}
\end{eqnarray}
A comment on the permissible values for the parameter $N$ is in
order. In the return distributions~(\ref{wisret})
and~(\ref{wisretmarginal}), $N$ can take any positive real value. In
the ensemble distribution~(\ref{wisal}), however, $N$ is the length of
the model time series or, equivalently, one of the dimensions of the
$K\times N$ matrices $A$ and thus restricted to integer values. It is
thus a matter of interpretation whether one wants to impose the
constraint that $N$ be integer. There is no such restriction for the
parameter $L$. In any case, we also carried out fits with the integer
constraint. The results shown in Fig.~\ref{fig3} do not indicate a
strong influence of this constraint. 

\begin{figure}[htbp]
  \begin{center}
    \includegraphics[width=0.4\textwidth]{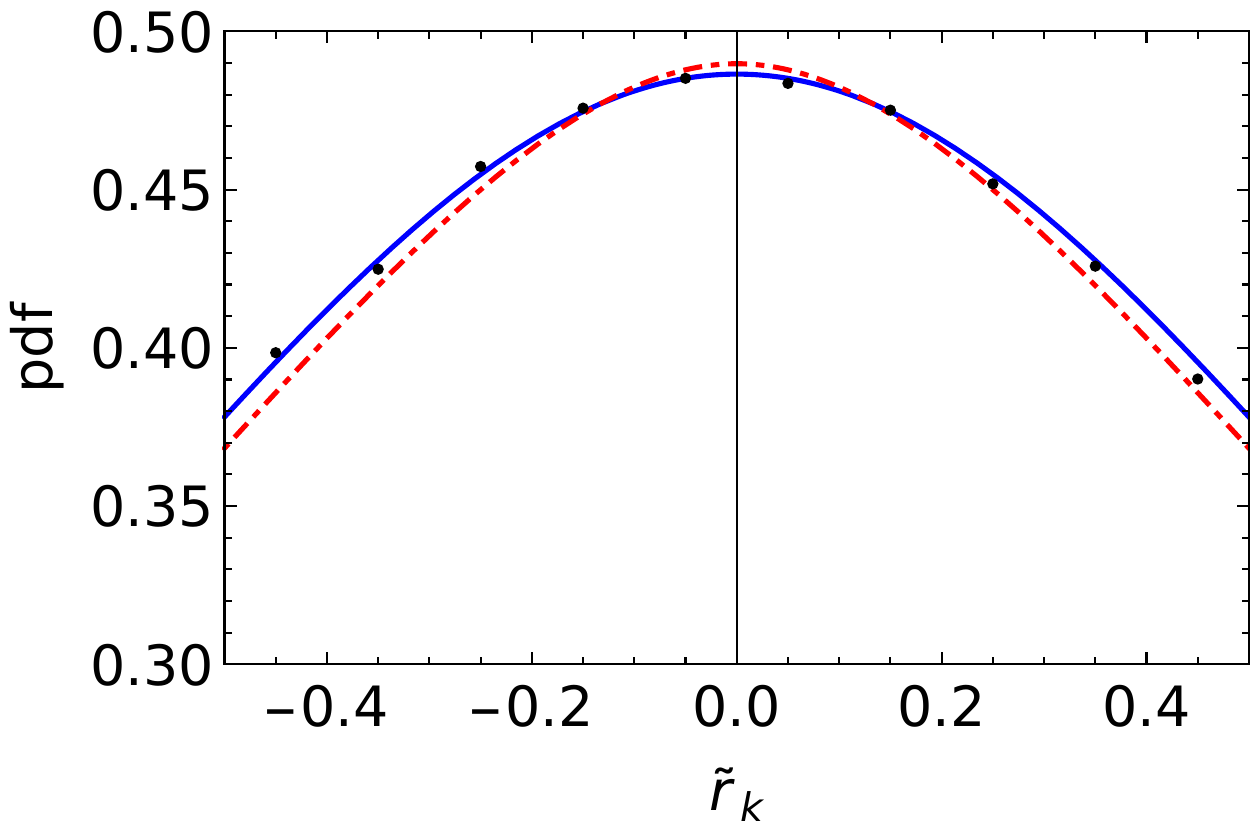}
  \end{center}
  \begin{center}
    \includegraphics[width=0.4\textwidth]{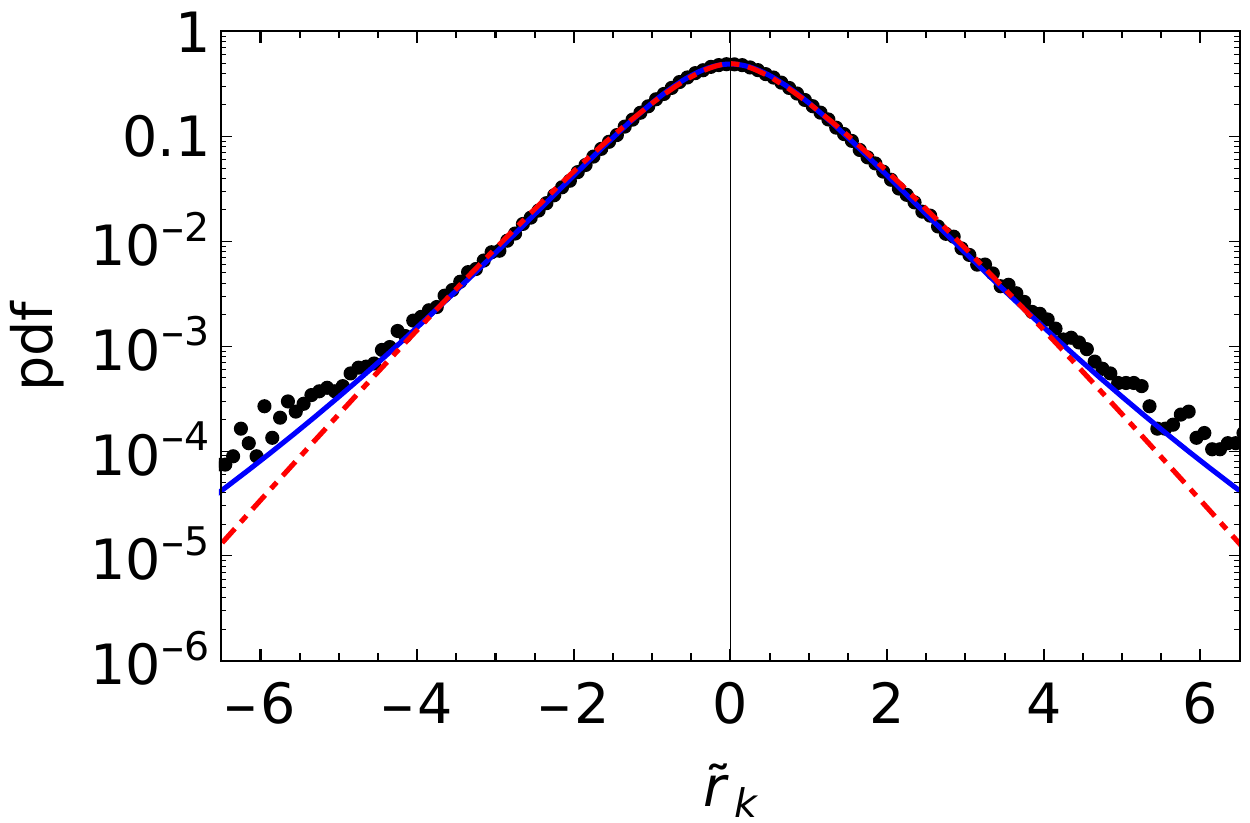}
  \end{center}
  \caption{Aggregated distribution of daily returns, $\Delta t=1$
    trading day. Empirical results as dots, fit to the
    distribution~(\ref{wisretmarginal}) as solid line. The
    corresponding result of Ref.~\cite{Schmitt2013} as dashed
    line. Center of the distribution on a linear scale (top), whole
    distribution on a logarithmic scale (bottom).}
 \label{fig4}
\end{figure}

\begin{figure}[htbp]
  \begin{center}
    \includegraphics[width=0.4\textwidth]{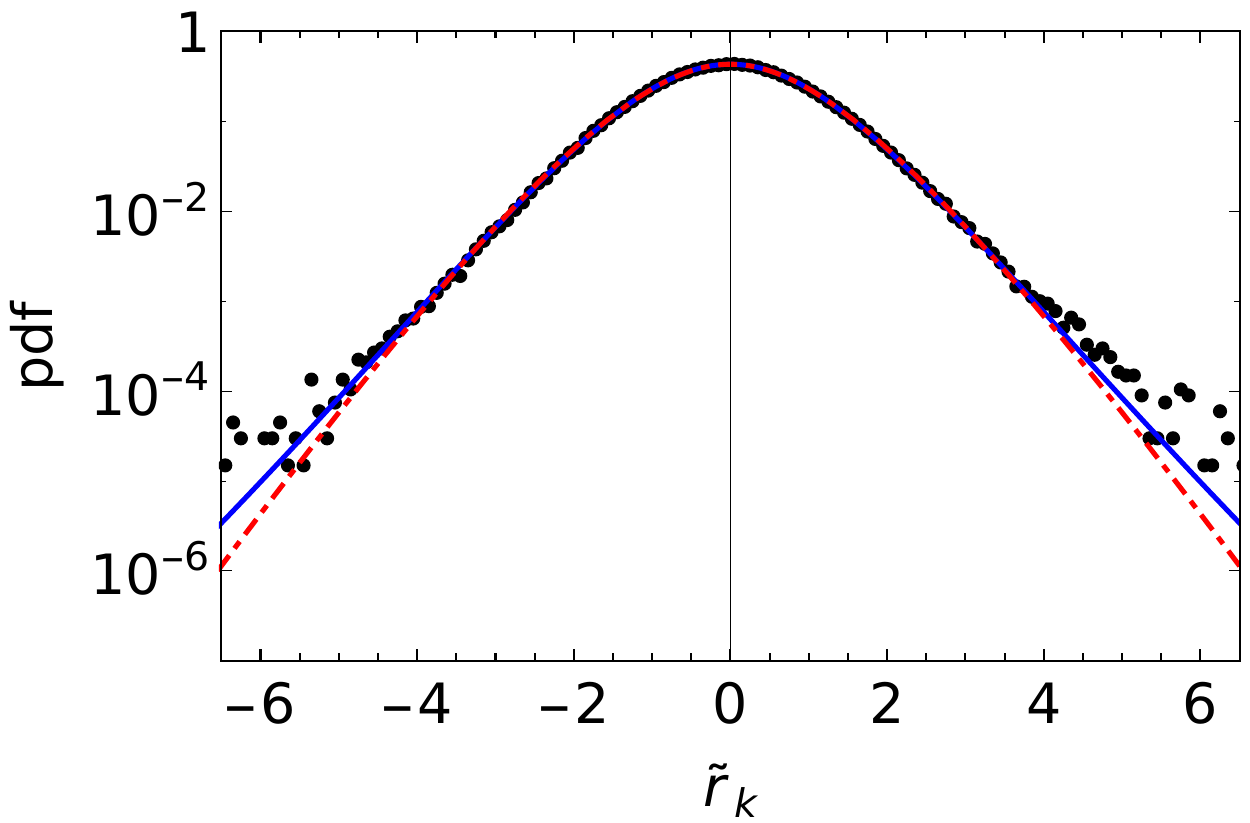}
  \end{center}
  \caption{Same as Fig.~\ref{fig4} (bottom) for returns with $\Delta t=20$
    trading days.}
 \label{fig5}
\end{figure}

The results of the data comparison are displayed in Fig.~\ref{fig4}
for daily returns, $\Delta t=1$ trading day. The fitted parameter values
are $N=8.13$ and $L=2.24$. The center of the empirical
distribution is slightly better described by employing the deformed
Wishart ensemble instead of the non--deformed one in
Ref.~\cite{Schmitt2013}. The heavy tails clearly reveal that the
deformed Wishart ensemble yields overall a better description, since
the result of Ref.~\cite{Schmitt2013} consistently underestimates the
large events. In Fig.~\ref{fig5} we present the same analysis for
returns with $\Delta t=20$ trading days, the fit
gives $N=20.98$ and $L=2.07$. Here, the tails are still strong, but
less pronounced than for daily data. For the interpretation of these
results, we recall the well--established fact that \textit{univariate}
distributions of returns for one stock acquire heavy tails as the
return horizon $\Delta t$ becomes smaller, see \textit{e.g.}
Ref.~\cite{doi:10.1080/713665670}.  Here, however, we analyze the
\textit{multivariate} distribution of $K$ correlated stocks. Thus,
there are two competing effects. First, as discussed in general in
connection with Eq.~(\ref{gaussian}) and for the financial data in
Ref.~\cite{Schmitt2013}, the superposition of the amplitudes, in the
present case the returns, drives the multivariate distribution towards
a Gaussian, provided that the covariances are sufficiently
constant. Second, as observed in Ref.~\cite{Schmitt2013} and extended
here, the fluctuations of the non--stationary covariances lift the
tails of the distributions evaluated over long time intervals and make
them heavier. Not surprisingly, the heavier the tails of the
\textit{univariate} distributions, the heavier are also those of
the ensemble averaged multivariate ones shown above. This is nicely
reflected in the nearly linear increase of the parameter $N$ on the
return horizon $\Delta t$, see Fig.~\ref{fig3}.  The smaller $N$, the
heavier are the tails in the ensemble distributions~(\ref{wisal}) and
in the enesmble averaged return distribution~(\ref{wisret}). As $N$
grows and $L$ is held fixed, the distribution~(\ref{wisal}) comes
closer to a Gaussian, \textit{i.e.}, to the non--deformed ensemble.
It is quite remarkable that the fit of $L$ always yields values close
to two. According to Eq.~(\ref{covard}), this implies
$\Sigma^{(d)}\approx\Sigma$. Put differently, the heavy tails in the
cases considered alter the measured covariances only slightly as
compared to a Gaussian assumption. When looking at financial risk,
however, the tails are very important.

\begin{figure}[htbp]
  \begin{center}
    \includegraphics[width=0.4\textwidth]{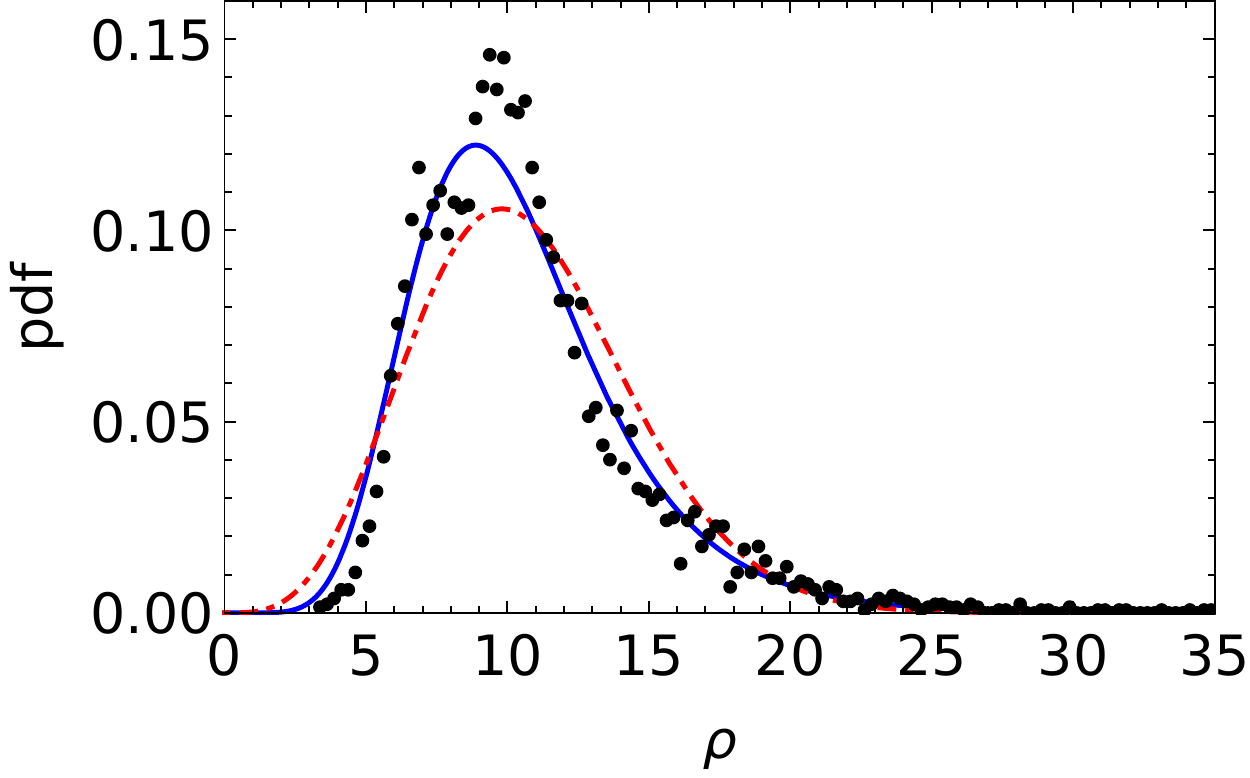}
  \end{center}
  \begin{center}
    \includegraphics[width=0.4\textwidth]{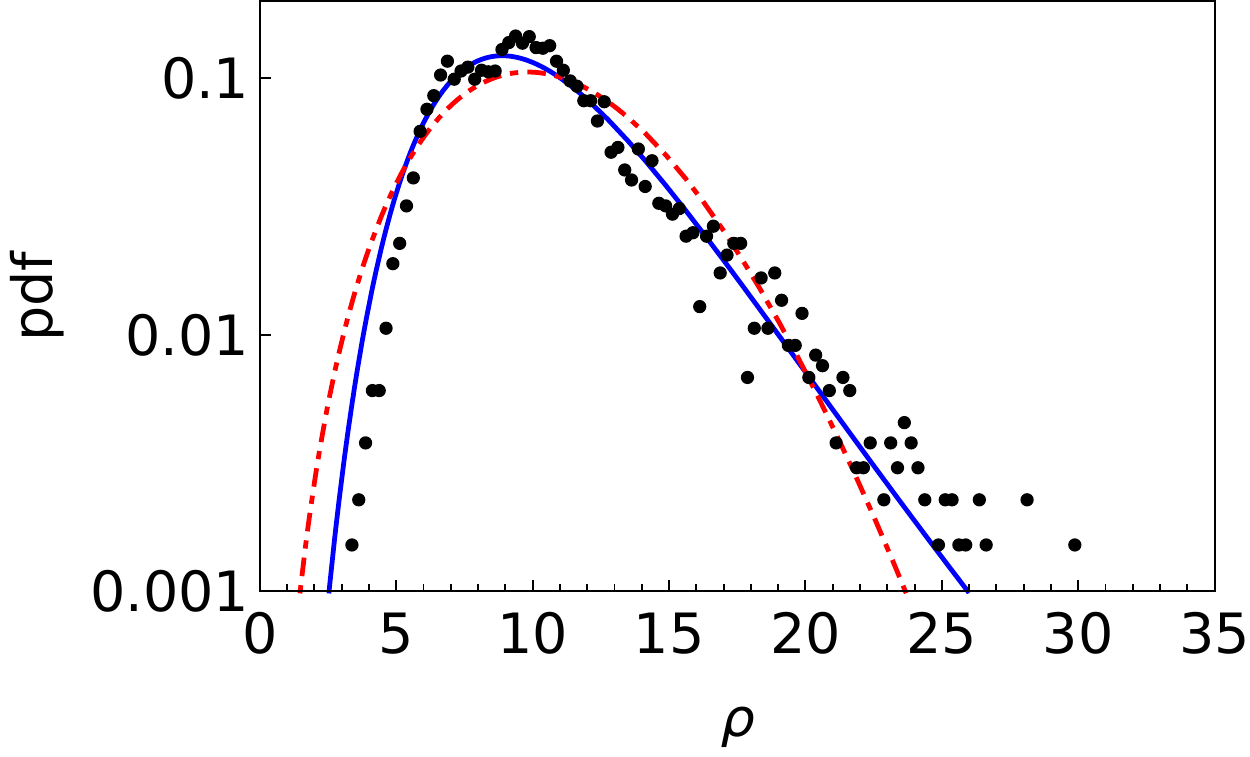}
  \end{center}
  \caption{Radial distribution of daily returns, $\Delta t=1$ trading
    day. Empirical results as dots, fit to the
    distribution~(\ref{disrad}) as solid line. The corresponding
    result of Ref.~\cite{Schmitt2013} as dashed line, on a linear
    (top) and a logarithmic scale (bottom).}
 \label{fig6}
\end{figure}

Finally, we use the opportunity to discuss an issue of general
interest when presenting and fitting a multivariate distribution that
depends on the statistical variables only via a bilinear form such as
$r^\dagger\Sigma^{-1}r$. Instead of the above procedure which involves
rotation of $r$ into the eigenbasis of $\Sigma$ and aggregation, one
might also view the bilinear form as a generalized radius
\begin{equation}
\rho = \sqrt{r^\dagger\Sigma^{-1}r}
\label{rho1}
\end{equation}
in the $K$ dimensional space and study its distribution 
\begin{equation}
\langle g_\textrm{rad}\rangle (\rho) = 
 \int \delta\left(\rho-\sqrt{r^\dagger\Sigma^{-1}r}\right) 
         \langle g\rangle (r|\Sigma,N,L) d[r] \ .
\label{rho2}
\end{equation}
A rather straightforward calculation yields
\begin{eqnarray}
& &\langle g_\textrm{rad}\rangle (\rho) = \frac{\Gamma(N+L/2)\Gamma((N+K+L)/2)}
                                    {2^{K/2-1}\Gamma(N/2)\Gamma(K/2)\Gamma((N+L)/2)} \nonumber\\
& & \qquad \rho^{K-1} \mathcal{U}\left(\frac{N+K+L}{2},\frac{K-N+2}{2},\frac{\rho^2}{2}\right) \ .
\label{disrad}
\end{eqnarray}
The Jacobian $\rho^{K-1}$, typical for such a radial distribution, appears
and, because of $K=306$, dominates the functional form of the
distribution for small $\rho$, as can be seen in Fig.~\ref{fig6}.
The theoretical result~(\ref{disrad}) describes the data much better
than the corresponding result of
Ref.~\cite{Schmitt2013}. Nevertheless, the dominance of $\rho^{K-1}$
gives a somewhat misleading picture and we infer that using the
distributions~(\ref{wisret}) and~(\ref{wisretmarginal}) is more
appropriate.

\section{Permissibility of Deformation Functions}
\label{sec4}

When extracting the return distribution deformation function $p(x)$
from the data, we encountered a puzzling problem that we wish to
report here. The log--logistic distribution
\begin{equation}
  p(x|b,c) = \frac{b}{c} \frac{(x/c)^{b-1}}{(1+(x/c)^b)^2}
 \label{logp}
\end{equation}
with $b=N/2$ yields a very good description of the data, even slightly
better than the beta prime distribution. The $c$ values are around
one, $N=4$ for $\Delta t=1$ trading day and increasing for larger
$\Delta t$. However, the resulting ensemble deformation function can
take positive and negative values. For example, for $N=4$, we find
\begin{equation}
  f(\eta|c) \sim \frac{\sin(\eta/2)-\eta\cos(\eta/2)/2}
                          {\eta^2} \ .
 \label{logf}
\end{equation}
Hence, it cannot be interpreted as a distribution. Indeed, we are
confronted with a problem of interpretation.  While the discussion in
Sec.~\ref{sec23} clearly revealed that $p(x)$ is a well--defined
distribution of a random variable, namely of a variance, it is not
obvious that $f(\eta)$ also represents a well--defined
distribution. The corresponding random variables, the matrices $A$, do
not have a direct data interpretation. Thus, one might simply view
$f(\eta)$ as a continuous coefficient function for the expansion of
the distribution~(\ref{defwis}) in terms of Gaussians. Nevertheless,
even if one does not want to enforce an interpretation of $f(\eta)$ as
a distribution, the resulting ensemble distribution~(\ref{defwis})
must be positive definite. We test this by calculating the
distribution of the traces
\begin{equation}
u(s) = \int \delta\left(s-\frac{1}{N}\Tr A^\dagger\Sigma^{-1}A\right)\overline{w}(A|N,c) d[A] \ .
\label{tracesig}
\end{equation}
As we only wish to test the positivity, it is convenient to choose it
different from the above distribution~(\ref{trace}) by including the
covariance matrix. After some algebra, we can express it as a
high--order derivative involving the return distribution deformation
function,
\begin{eqnarray}
u(s) &=& \frac{(-1)^{(K-1)N/2}\Gamma(N/2)}{\Gamma(KN/2)}s^{KN/2-1} \nonumber\\
     & & \qquad\qquad \frac{d^{(K-1)N/2}}{ds^{(K-1)N/2}}\frac{p(s|b,c)}{s^{N/2-1}} \ .
\label{u1}
\end{eqnarray}
This in principle general result gives for the log--logistic
distribution~(\ref{logp})
\begin{eqnarray}
u(s) &=& \frac{(-1)^{(K-1)N/2}Nc^{N/2}\Gamma(N/2)}{2\Gamma(KN/2)}s^{KN/2-1} \nonumber\\
     & & \qquad\qquad  \frac{d^{(K-1)N/2}}{ds^{(K-1)N/2}}\frac{1}{(c^{N/2}+s^{N/2})^2} \ .
\label{u2}
\end{eqnarray}
Restricting ourselves to even $N$, we may employ the theory of complex
functions to calculate the pole expansion
\begin{eqnarray}
& &\frac{1}{(c^{N/2}+s^{N/2})^2} = \sum_{n=1}^{N/2} \frac{1}{\prod_{m\neq n} (a_n-a_m)^2}\nonumber\\
& & \qquad\qquad \left(\frac{1}{(s-a_n)^2}-\frac{2}{s-a_n}\sum_{l\neq n}\frac{1}{a_n-a_l}\right)
\label{poleexp}
\end{eqnarray}
with the poles
\begin{equation}
a_n = c \exp\left(\frac{i2\pi}{N}(2n+1)\right) \ .
\label{poles}
\end{equation}
The derivatives in Eq.~(\ref{u2}) can now easily be evaluated and we
arrive at
\begin{eqnarray}
u(s) &=& \frac{Nc^{N/2}\Gamma(N/2)}{2\Gamma(KN/2)}s^{KN/2-1}\nonumber\\
      & & \quad 
               \sum_{n=1}^{N/2} \frac{1}{\prod_{m\neq n} (a_n-a_m)^2}
             \Biggl(\frac{\Gamma((K-1)N/2+2)}{(s-a_n)^{(K-1)N/2+2}}\nonumber\\
      & & \qquad -
    \frac{2\Gamma((K-1)N/2+1)}{(s-a_n)^{(K-1)N/2+1}}\sum_{l\neq n}\frac{1}{a_n-a_l}\Biggr) \ .
\label{u3}
\end{eqnarray}
Inspite of the complex poles, this is by construction a real function.
Yet, it takes positive and negative values which outrules an
interpretation of $u(s)$ and thus also of $\overline{w}(A|N,c)$ as
distributions. By means of this example we face the somewhat
surprising result that a well--defined distribution $p(x)$ does not
necessarily yield a well--defined ensemble. Each case has to be
investigated individually.

\section{Further Extension by Deforming the Static Amplitude
                 Distribution}
\label{sec5}

We argued in Sec.~\ref{sec21} that the Gaussian
assumption~(\ref{gaussian}) for the static amplitude distribution is
not as restrictive as it might appear at first sight. Nevertheless, we
now extend our construction by assuming more general functional
forms. At present, we do not have data at our disposal in which the
static amplitude distribution is non--Gaussian, but we nevertheless
now extend our construction, as it might be useful for future data
analyses.  Moreover, we will also come across some interesting
observations.  Instead of Eq.~(\ref{gaussian}), we now assume that the
static amplitude distribution can be expressed as an average over the
Gaussian~(\ref{gaussian}),
\begin{equation}
\overline{g}(r|\Sigma_s) = \int\limits_0^\infty h(\xi) 
                 g\left(r\left|\frac{\Sigma_s}{\xi}\right.\right) d\xi
\label{static}
\end{equation}
with a new deformation function $h(\xi)$ that fulfils
\begin{align}
\int\limits_0^\infty h(\xi) d\xi = 1 \quad \textrm{and} \quad h(\xi) \ge 0 \ .
\label{hconstraints}
\end{align}
We proceed as in Sec.~\ref{sec22}. Instead of the ensemble
average~(\ref{average}), we now have
\begin{equation}
 \langle\overline{g}\rangle (r|\Sigma,N) = \int d[A] \overline{w}(A|\Sigma,N) 
                    \overline{g}\left(r\left|\frac{1}{N}AA^\dagger\right.\right) \ .
\label{averageg}
\end{equation}
This is readily cast into the form
\begin{equation}
 \langle\overline{g}\rangle (r|\Sigma,N) = \int\limits_0^{\infty} p(x)
                    g\left(r|x\Sigma\right) dx \ ,
\label{averagesingleg}
\end{equation}
which differs from Eq.~(\ref{averagesingle}) only by the definition of
the amplitude distribution deformation function. It is now given by
\begin{equation}
p(x) = \int\limits_0^{\infty}d\xi h(\xi) \int\limits_0^{\infty}d\eta f(\eta) 
              \int\limits_0^\infty dz \chi^2_N(z) \delta\left(x-\frac{z}{\xi\eta}\right) \ .
\label{aampldefg}
\end{equation}
For fixed $\xi$, we introduce the new variable $\hat{\eta}=\eta\xi$ and find
\begin{equation}
p(x) = \int\limits_0^{\infty}d\hat{\eta} \hat{f}(\hat{\eta}) 
              \int\limits_0^\infty dz \chi^2_N(z) \delta\left(x-\frac{z}{\hat{\eta}}\right) \ .
\label{aampldefg2}
\end{equation}
which coincides with Eq.~(\ref{aampldef}), but now involving the new
ensemble deformation function
\begin{equation}
\hat{f}(\hat{\eta}) = \int\limits_0^{\infty} \frac{h(\xi)}{\xi} 
                         f\left(\frac{\hat{\eta}}{\xi}\right) d\xi \ .
\label{aampldefg3}
\end{equation}
This integral is reminiscent of a convolution.  Thus, the case of a
deformed, non--Gaussian static amplitude distribution is formally
traced back to the Gaussian case. The difference can be fully absorbed
into the ensemble deformation function. Importantly, this means that
all other results of Sec.~\ref{sec2} continue to hold, in particular
the Laplace transform~(\ref{laplace}) and its
inversion~(\ref{invlaplace}). Nevertheless, the following problem
remains. We can extract $h(\xi)$ and $p(x)$ from the data by using the
methods outlined in Sec.~\ref{sec23} for very short time intervals and
for the whole, long time interval, respectively. From the inverse
Laplace transform~(\ref{invlaplace}), we obtain $\hat{f}(\hat{\eta})$,
but to determine $f(\eta)$, we are left with the task to invert
Eq.~(\ref{aampldefg3}). Although that is definitely possible for some
special cases, a general inversion formula is lacking. In practical
applications, however, the extension sketched above is more likely to
be needed for consistency tests. For example, if some of the available
data for the same system permit the Gaussian assumption for the static
amplitude distribution and others do not, one can first determine
$f(\eta)$ as described in Sec.~\ref{sec2} and then turn to the data
which require an additional deformation function $h(\xi)$. Once both
of these deformation functions are known, one can evaluate
$\hat{f}(\hat{\eta})$ and check if it is consistent with the
inverse~(\ref{invlaplace}) of $p(x)$ which is independently extracted
from the data.

As an example, we consider the case that both, $f(\eta)$ and $h(\xi)$,
are $\chi^2$ distributions
\begin{equation}
f(\eta) = \chi_{N+L}^2(\eta) \qquad \textrm{and} \qquad 
h(\xi) = \chi_M^{2}(\xi)
\label{aampldefg4}
\end{equation}
of $N+L$ and $M$ degrees of freedom, respectively. The choice for
$f(\eta)$ coincides with the result of Sec.~\ref{sec32}. With
Eq.~(\ref{aampldefg3}), we obtain
\begin{eqnarray}
\hat{f}(\hat{\eta}) = \frac{\sqrt{\hat{\eta}}^{(N+L+M)/2-2}\mathcal{K}_{(N+L-M)/2}(\sqrt{\hat{\eta}})}
                              {2^{(N+L+M)/2-1}\Gamma((N+L)/2)\Gamma(M/2)} \ ,
\label{wisretext}
\end{eqnarray}
where $\mathcal{K}_\nu$ is the modified Bessel function of the second
kind of order $\nu$. This function already appeared in the ensemble
averaged return distribution of Ref.~\cite{Schmitt2013}. According to
Eq.~(\ref{aampldefg2}), the distribution $p(x)$ is an integral
involving the modified Bessel function and the return distribution
averaged over the deformed ensemble is an integral over a product of 
modified Bessel functions, but we do not give the formulae here.

\section{Conclusions}
\label{sec6}

Non--stationarity is an often encountered feature in complex systems.
Here, we addressed non--stationarity of correlations. We presented a
method to determine their distribution from the amplitude
distribution. Put differently, we showed how to extract the proper
ensemble of random covariance matrices from amplitude data.  Carrying
out our analysis for the case of financial data, we found an algebraic
distribution of covariance matrices reflecting the heavy tails in the
amplitude, \textit{i.e.}, return distributions.

A conceptually important comment is in order. Consider any two
empirical covariance matrices $\Sigma(t_1)$ and $\Sigma(t_2)$.  They
are certainly not independent, because, first, the correlation
structure due to, \textit{e.g.}, the industrial sectors in the
financial markets only changes on a large time scale and, second, one
would expect that they are the more dependent the smaller the time
difference $|t_1-t_2|$. The first cause is a built--in feature of our
model, as the random model covariance matrices fluctuate around the
empirical $\Sigma$. The second cause is effectively accounted for as
the length of our model time series $N$, \textit{i.e.}, the second
dimension of the random matrices $A$, is different from the length $T$
of the subintervals in which the empirical covariance matrices
$\Sigma(t)$ are evaluated. The values for $N$ resulting from the data
analysis are very small $N\ll K$ while $T\ge K$ when measuring
$\Sigma(t)$.  Our random matrix ansatz does not aim at modeling the
ensemble of the empirical covariance matrices $\Sigma(t)$ in a
one--to--one fashion. This is never the goal of a statistical
approach. Our model has an effective character which is obvious, for
example, in the relation $N\ll T$. Model time series much shorter than
the empirical ones suffice to properly grasp the statistical effects
induced by the fluctuations around the average covariance matrix. This
also reflects the mutual dependence of the empirical covariance
matrices. We demonstrated by our analysis of financial data how useful
our model is.

Furthermore, observables such as the amplitude distribution do not
resolve autocorrelations in time of any kind. Even reshuffeling the
order of the empirical covariance matrices $\Sigma(t)$ in time does
not change the amplitude distribution for the total time interval. A
similar situation is encountered when analyzing statistical properties
of quantum chromodynamics. The gauge fields may be viewed as a really
existing ensemble over which an average is carried out --- this is the
very definition of the partition function. Although these gauge fields
are not independent either, an effect of this autocorrelation is only
seen if corresponding observables are used. Densities are not
affected.  Random matrices as models for the highly non--trivial gauge
fields are applied with great success.

Yet another aspect is worth mentioning. In contrast to random matrix
applications for Hamiltonian systems, there is not a local scale that
can enforce universal statistical behavior.  Thus, it is important how
the covariances are actually distributed.  Gaussian assumptions are
only acceptable if really justified by data analysis. The present
study extends a previous one in which we employed such a Gaussian
assumption in finance. Here, we reconsidered the same data set and
clearly demonstrated that the Gaussian assumption underestimates the
tails. The algebraic distributions that we found here are relevant for
risk estimation as they help to better understand large
events. Importantly, once the ensemble is properly extracted,
meaningful averages can be computed for all observables that depend on
non--stationary covariances.

When developing our construction, we came across a puzzling feature
which calls for a caveat. The deformation function extracted from the
amplitude distribution determines, on the one hand, uniquely the
ensemble, but, on the other hand, this ensemble is not necessarily
well--defined. Each case has to be studied individually. We do not expect
this to cause severe problems in applications, but conceptually it is
an interesting aspect. We also extended our construction by including
deformed static amplitude distributions. The additional freedom
accompanying this extension might offer a possibility to circumvent
the above mentioned puzzling problem. From a more general viewpoint,
we have to emphasize that our construction only includes functional
forms of the ensemble that depend on the trace over the product of the
random covariance matrix and the mean covariance matrix. Although this
is quite natural, as it guarantees a certain amount of basis
invariance which all random matrix models need, more general
functional forms pose an interesting and potentially important
challenge.

Hitherto, we only applied our method to finance. We plan applications
to other complex systems, too. This may be rewarding, as large events
and risk estimations are not only important in finance.

\section*{Acknowledgments}

We thank Desislava Chetalova, Tobias Nitschke, Thilo Schmitt and Yurij
Stepanov for fruitful discussions.


\end{document}